\documentclass[
a4paper, 12pt, ,nofootinbib, groupedaddress, superscriptaddress, amsmath, amssymb]{revtex4}
\usepackage{graphicx}
\usepackage{dcolumn}
\usepackage{bm}
\usepackage{amssymb}
\usepackage{amsmath}
\usepackage{epsfig}    
\usepackage{color}
\usepackage{slashed}
\usepackage{hhline}
\def\be{\begin{equation}}
\def\ee{\end{equation}}
\newcommand{\bea}{\begin{eqnarray}}
\newcommand{\eea}{\end{eqnarray}}
\newcommand{\nn}{\nonumber}

\numberwithin{equation}{section}

\begin{document}

{\begin{flushright}{CTP-SCU/2021014, APCTP Pre2021-08}
\end{flushright}}

\title{Explanations for anomalies of  muon anomalous magnetic dipole moment, $b\to s \mu\bar\mu$ and radiative neutrino masses in a leptoquark model
}

\author{Takaaki Nomura}
\email{nomura@scu.edu.cn}
\affiliation{College of Physics, Sichuan University, Chengdu 610065, China}

%
\author{Hiroshi Okada}
\email{hiroshi.okada@apctp.org}
\affiliation{Asia Pacific Center for Theoretical Physics (APCTP) - Headquarters San 31, Hyoja-dong,
  Nam-gu, Pohang 790-784, Korea}
\affiliation{Department of Physics, Pohang University of Science and Technology, Pohang 37673, Republic of Korea}

\date{\today}

\begin{abstract}
We propose a leptoquark model simultaneously to explain anomalies of muon anomalous magnetic dipole moment and $b\to s \mu\bar\mu$ in light of experimental reports very recently. 
Here, we satisfy several stringent constraints such as $\mu\to e\gamma$ and meson mixings.
In addition, we find these leptoquarks also play an role in generating tiny neutrino masses at one-loop level without introducing any additional symmetries. We have numerical analysis and show how degrees our parameter space is restricted.
\end{abstract}

\maketitle
\newpage
\section{Introduction}

The anomalous magnetic dipole moment of muon (muon $g-2$) is precisely predicted in the SM and the deviation from the prediction indicates new physics beyond the SM. 
The E821 experiment at Brookhaven National Lab (BNL) reported a $3.3 \sigma$ deviation from the SM prediction that is written by~\cite{Zyla:2020zbs, Bennett:2006fi,Davier:2019can}
\begin{equation}
\Delta a_\mu  = a^{\rm exp}_\mu - a^{\rm SM}_\mu = (26.1 \pm 7.9) \times 10^{-10}.
\end{equation}
Moreover a $3.7 \sigma$ deviation was given by applying the lattice calculations as $\Delta a_\mu = (27.4 \pm 7.3) \times 10^{-10}$ in ref.~\cite{Blum:2018mom} and $\Delta a_\mu = (27.06 \pm 7.26) \times 10^{-10}$ in ref.~\cite{Keshavarzi:2018mgv}.
Thus muon $g-2$ is a long-standing anomaly in particle physics and various solutions have been considered; review can be found in~\cite{Jegerlehner:2009ry,Miller:2012opa,Lindner:2016bgg,Jegerlehner:2018zrj}.
Very recently the new muon $g-2$ measurement  in E989 experiment at Fermilab reported the new result indicating~\cite{Abi:2021gix} 
\begin{equation}
a_\mu^{\rm FNAL} = 116592040(54) \times 10^{-11}.
\end{equation}
Combining BNL result we have new $\Delta a_\mu$ value of
\begin{equation}
\label{eq:amu_new}
\Delta a_\mu^{\rm new} = (25.1 \pm 5.9) \times 10^{-10} .
\end{equation}
The deviation from the SM is now 4.2 $\sigma$.

Experimental anomalies in rare $B$ meson decays ($b \to s \ell \bar\ell$) also indicate deviation from the SM prediction.
There are discrepancies in the measurements of the angular observable $P'_5$ in the $B$ meson decay 
($B \to K^* \mu^+ \mu^-$)~\cite{DescotesGenon:2012zf, Aaij:2015oid, Aaij:2013qta,Abdesselam:2016llu, Wehle:2016yoi}, 
the lepton non-universality indicated by the ratio of branching fractions, $R_K = BR(B^+ \to K^+ \mu^+\mu^-)/ BR(B^+ \to K^+ e^+e^-)$~\cite{Hiller:2003js, Bobeth:2007dw, Aaij:2014ora,Aaij:2019wad}, 
and $R_{K^*} = BR (B \to K^* \mu^+ \mu^-)/ BR (B \to K^* e^+ e^-)$~\cite{Aaij:2017vbb}.
Recently, the LHCb collaboration reported the updated result
of $R_{K}$ as~\cite{Aaij:2021vac}  
\begin{equation}
R_{K} = 0.846^{+ 0.042 + 0.013}_{- 0.039 -0.012} \quad (1.1 {\rm GeV}^2 < q^2 < 6 {\rm GeV}^2),
\end{equation}
where first(second) uncertainty is statical(systematic) one and $q^2$ is the invariant mass squared for dilepton.
Remarkably the central value is the same as previous result~\cite{Aaij:2019wad} and we have $3.1 \sigma$ deviation from the SM prediction. 
Motivated by these results, various global analysis for relevant Wilson coefficients are also carried out~\cite{Ciuchini:2019usw,Descotes-Genon:2015uva,Alguero:2019ptt,Aebischer:2019mlg,Alok:2019ufo}, 
which indicate negative contribution to Wilson coefficients $C_9^\mu$ and $C_{10}^\mu$ associated with $(\bar s_R \gamma^\mu b_L)( \bar \mu \gamma_\mu \mu)$ and  $(\bar s_R \gamma^\mu b_L)( \bar \mu \gamma_\mu \gamma_5 \mu)$ operators; global analyses with recent results including $B_s \to \mu^+\mu^-$ result by LHCb~\cite{Aaij:2021vac} is found in ref.~\cite{Altmannshofer:2021qrr,Kriewald:2021hfc,Cornella:2021sby,Alok:2019ufo}.

{
To explain the flavor issues above, one of the attractive scenarios is to introduce leptoquarks since they can interact with both leptons and quarks.
In fact it is possible to explain muon $g-2$ and/or $b \to s \ell \bar\ell$ anomalies by various leptoquark models~\cite{Chen:2017hir,Chen:2016dip,Babu:2020hun,Crivellin:2020tsz,Datta:2019bzu,Popov:2019tyc,Cheung:2016fjo,Hiller:2016kry,Bauer:2015knc,Sahoo:2015fla,ColuccioLeskow:2016dox,Becirevic:2016oho,Becirevic:2016yqi,Sahoo:2016pet,Cheung:2017efc,Cai:2017wry,Sahoo:2015wya,Popov:2016fzr,Crivellin:2019dwb,Greljo:2021xmg,Angelescu:2021lln,Dorsner:2019itg,Arnan:2019olv}.
We can obtain contribution to the Wilson coefficients $C_{9,10}^\mu$ for explaining $ b \to s \ell \bar \ell $ anomalies at tree level by introducing scalar leptoquarks with charge assignments $({\bf 3}, {\bf 2}, 7/6)$ and 
$(\bar {\bf 3}, {\bf 3}, 1/3)$  under (SU(3), SU(2), U(1))~\cite{Sahoo:2015wya, Chen:2016dip, Becirevic:2016oho, Cheung:2016fjo, ColuccioLeskow:2016dox, Chen:2017hir, Arnan:2019olv, Popov:2019tyc,Crivellin:2019dwb,Crivellin:2020tsz, Babu:2020hun}. 
In particular we can obtain different magnitude of $C_9^\mu$ and $C_{10}^\mu$ by combining these leptoquarks improving fit to the experimental data~\cite{Chen:2016dip, Chen:2017hir}.
Moreover the leptoquark with charge $({\bf 3}, {\bf 2}, 7/6)$ can provide sizable muon $g-2$ since it can couples to both left- and right-handed muon avoiding chiral suppression~\cite{ColuccioLeskow:2016dox, Chen:2017hir,Crivellin:2020tsz,Babu:2020hun}.
Yukawa interactions associated with some scalar leptoquarks can also generate active neutrino masses at loop level~\cite{Cheung:2016fjo,Babu:2020hun,Cheung:2017efc,Cai:2017wry,Popov:2016fzr}.
Remarkably we can realize neutrino mass by adding only two leptoquarks with charge assignments  $({\bf 3}, {\bf 2}, 1/6)$ and $(\bar{\bf 3}, {\bf 3}, 1/3)$ where we do not need any extra symmetry such as $Z_2$~\cite{Cheung:2016fjo};  
note that other extra contents such as extra scalar fields and/or vector-like fermions are required to generate radiative neutrino mass in other leptoquark combinations~\cite{Babu:2020hun,Cheung:2017efc,Cai:2017wry,Popov:2016fzr}.
It is thus interesting to combine leptoquarks with charges $({\bf 3}, {\bf 2}, 7/6)$, $(\bar{\bf 3}, {\bf 3}, 1/3)$ and $({\bf 3}, {\bf 2}, 1/6)$ to explain $b \to s \ell \bar \ell$ anomalies, muon $g-2$, and neutrino mass
~\footnote{In our previous work in ref.~\cite{Cheung:2016fjo} it is difficult to explain $b \to s \ell \bar \ell$ anomalies while fitting neutrino data and introduction of three leptoquarks is important.}.
}


In this work we propose a model with three scalar leptoquarks. 
Combination of interactions among these leptoquarks and the SM fermions can explain muon $g-2$, $b \to s \ell \ell$ anomaly and radiative neutrino mass generation at the same time.
For neutrino mass generation, we adopt the mechanism in ref.~\cite{Cheung:2016fjo} realized by two leptoquarks, and one $SU(2)_L$ doublet leptoquark is added to improve $b \to s \ell \bar\ell$ explanation 
and to realize sizable muon $g-2$. 
We then analyze our model to find a solution to explain these issues taking into account possible flavor constraints such as lepton flavor violating (LFV) decays of charged leptons ($\ell \to \ell' \gamma$) 
and mixing between meson and anti-meson ($M$--$\overline M$ mixing).

This paper is organized as follows.
In Sec.~II, we review our model and show relevant formulas for neutrino mass, Wilson coefficients for $b \to s \ell \bar\ell$ decay, LFV branching ratios(BRs), muon $g-2$ and $M$--$\overline M$ mixing for our analysis.
In Sec.~III, we carry out numerical analysis taking into account flavor constraints and show muon $g-2$ and Wilson coefficients which are compared with recent measurements (global fit results).
We conclude in Sec.~IV.

\section{Model setup and Constraints}
\begin{table}[tb]
\begin{tabular}{|c||c|c|c|}
\hline\hline  
                   & ~$\eta$~  & ~$\eta'$~  & ~$\Delta$ \\\hline 
$SU(3)_C$ & $\bm{3}$ & $\bm{3}$  & $\bar{\bm3}$  \\\hline 
$SU(2)_L$ & $\bm{2}$& $\bm{2}$  & $\bm{3}$  \\\hline 
$U(1)_Y$   & $\frac16$  & $\frac76$ & $\frac13$    \\\hline
\end{tabular}
\caption{\small 
Charge assignments of the leptoquark fields $\eta$, $\eta'$ and $\Delta$  
under $SU(3)_C\times SU(2)_L\times U(1)_Y$.}
\label{tab:1}
\end{table}

We review our model setup in this section.
We introduce three types of leptoquark bosons $\eta$, $\eta'$, and $\Delta$.
$\eta$ has triplet under $SU(3)_C$, doublet under $SU(2)_L$, and $1/6$ under $U(1)_Y$, and dominantly contributes to the neutrino masses.
  $\Delta$ has anti-triplet under $SU(3)_C$, triplet under $SU(2)_L$, and $1/3$ under $U(1)_Y$, and mainly contributes to the neutrino masses together with $\eta$,
lepton anomalous magnetic moment, and lepton universality of $b\to s\mu\bar\mu$  both of which are recently reported that there could exist anomalies beyond the SM. 
$\eta'$ has the same charges as $\eta$ under $SU(3)_C$ and $SU(2)_L$, but $7/6$ under $U(1)_Y$, and dominantly plays an role in suppressing the experimental bounds of $B\to \mu\bar\mu$
and BR($\mu\to e\gamma$).

The new field contents and their charges are shown in Table~\ref{tab:1}.
The relevant Lagrangian for the interactions of $\eta$ and $\Delta$ with
fermions and the Higgs field is given by 
\begin{align}
-\mathcal{L}_{Y}
&=
 f_{ij} \overline{d_{R_i}} \tilde \eta^\dag L_{L_j} + g_{ij} \overline{ Q^c_{L_i}}
 (i\sigma_2) \Delta L_{L_j} 
+h_{ij} \bar{Q}_{L_i} \eta' \ell_{Rj} + \tilde{h}_{ij} \bar L_{L_i} \tilde \eta' u_{Rj} 
+ {\rm h.c.},\\
{\cal V}_{non-tri} & =  \mu H^\dag \Delta \eta + \lambda_0 (H^T H)_{\bf 3} [\eta'^\dag \tilde\eta^*]_{\bf 3}+ {\rm h.c.},
\label{Eq:lag-flavor}
\end{align}
where $(i,j)=1-3$ are generation indices, 
$\tilde\eta\equiv i\sigma_2\eta^*$, $\sigma_2$ is the second 
Pauli matrix, and $H$ is the SM Higgs field that develops a 
nonzero vacuum expectation value (VEV), which is denoted by 
$\langle H\rangle\equiv [0,v/\sqrt2]^T$. 
{\it Although $\lambda_0$ induces the mixing between $\eta_{2/3}$ and $\eta'_{2/3}$, we neglect this term for simplicity.}
We work in the basis where all the coefficients are real and positive
for simplicity.~\footnote{If we consider this mixing, it affects the neutrino masses, LFVs, muon $g-2$.} 
The scalar fields can be parameterized as 
\begin{align}
&H =\left[
\begin{array}{c}
w^+\\
\frac{v+\phi+iz}{\sqrt2}
\end{array}\right],\quad 
\eta =\left[
\begin{array}{c}
\eta_{2/3}\\
\eta_{-1/3}
\end{array}\right],\quad 
\eta' =\left[
\begin{array}{c}
\eta'_{5/3}\\
\eta'_{2/3}
\end{array}\right],\quad 
\Delta =\left[
\begin{array}{cc}
\frac{\delta_{1/3}}{\sqrt2} & \delta_{4/3} \\
\delta_{-2/3} & -\frac{\delta_{1/3}}{\sqrt2}
\end{array}\right],
\label{component}
\end{align}
where the subscript of the fields represents the electric charge, 
$v \approx 246$ GeV, and $w^\pm$ and $z$ are, respectively, the 
Nambu-Goldstone bosons, which are absorbed by the 
longitudinal components of the $W$ and $Z$ bosons.
Due to the $\mu$ term in Eq.~(\ref{Eq:lag-flavor}), the charged components
with $1/3$ and $2/3$ electric charges mix each other, and 
their mixing matrices and mass eigenstates are defined as follows: 
\begin{align}
&\left[\begin{array}{c} \eta_{i/3} \\ \delta_{i/3} \end{array}\right] = 
O_i \left[\begin{array}{c} A_i \\ B_i \end{array}\right],\quad
O_i\equiv 
\left[\begin{array}{cc} c_{\alpha_i} & s_{\alpha_i} \\
 -s_{\alpha_i} & c_{\alpha_i}   \end{array}\right], \quad (i=1,2),
\end{align}
where their mass eigenstates are denoted as $m_{A_i}$ and  $m_{B_i}$, respectively. 
Then whole the interaction in terms of the mass eigenstates can be written by 
\begin{align}
 - {L}_{Y}\approx &
f_{ij} \overline{ d_{R_i}} \nu_{L_j} (c_{\alpha_1} A_1 +s_{\alpha_1} B_1)
-\frac{g_{ij}}{\sqrt2} \overline{ d_{L_i}^c} \nu_{L_j} (-s_{\alpha_1} A_1 + 
c_{\alpha_1} B_1) 
\nn\\
&-
f_{ij} \overline{d_{R_i}} \ell_{L_j} (c_{\alpha_2} A_2 +s_{\alpha_2} B_2)
-
\frac{g_{ij}}{\sqrt2} \overline{u_{L_i}^c} \ell_{L_j} (-s_{\alpha_1} A_1 + 
  c_{\alpha_1} B_1) 
 \nn\\
&
-
{g_{ij}} \overline{d_{L_i}^c} \ell_{L_j}\delta_{4/3} \;
 + \; 
  {g_{ij}} \overline{u_{L_i}^c} \nu_{L_j} (-s_{\alpha_2} A_{2}^* + c_{\alpha_2} B_{2}^*) \nonumber \\
&+h_{ij} \left[ \bar u_{Li} \, \ell_{Rj} \eta'_{5/3}  + \bar d_{Li} \, \ell_{R j} \eta'_{2/3}  \right] 
+  \tilde{h}_{ij} \left[ \bar \ell_{L i}  \, u_{Rj} \eta'_{-5/3} - \bar \nu_{L i}\, u_{R j} \eta'_{-2/3} \right]   .
\label{eq:lag}
\end{align}
In the following we summarize various phenomenological formulae derived from these interactions.

\subsection{Neutrino mixing}
In our model active neutrino mass can be generated at one loop level via interactions among leptons, quarks and leptoquarks 
where the mechanism is the same as the one in ref.~\cite{Cheung:2016fjo}.
Calculating one-loop diagram the  active neutrino mass matrix $m_\nu$
is given by
\begin{align}
&(m_{\nu})_{ab}
=
\sum_{i=1}^3
\left[g^T_{bi} R_i f_{ia}+ f_{ai} R_i g_{ib}^T \right] ,\quad
R_i
\approx
- \frac{3 s_{2\alpha_1}}{2(4\pi)^2}  \ln\left(\frac{m_{A_1}}{m_{B_1}}\right) m_{d_i} ,
\label{eq:R}
\end{align}
where we assume $m_{d_i}<< m_{A_1},m_{B_1}$.
$({m}_\nu)_{ab}$ is diagonalized by the Pontecorvo-Maki-Nakagawa-Sakata 
mixing matrix $V_{\rm MNS}$ (PMNS)~\cite{Maki:1962mu} as 
$
({m}_\nu)_{ab} =(V_{\rm MNS} D_\nu V_{\rm MNS}^T)_{ab}$ with $D_\nu\equiv 
(m_{\nu_1},m_{\nu_2},m_{\nu_3})$.
Then, we rewrite $g$ in terms of observables~\cite{Esteban:2020cvm} and several input parameters as follows~\cite{Nomura:2016pgg, Cheung:2016fjo}:
\begin{align}
\label{eq:g}
 g = \frac12  R^{-1} (f^T)^{-1} (V_{\rm MNS} D_\nu V_{\rm MNS}^T+ A),
\end{align}
where $A$ is an arbitrary three by three antisymmetric matrix with complex values,
and perturbative limit $g\lesssim \sqrt{4\pi}$ has to be satisfied.

\subsection{$b\to s\ell\bar\ell$}
In our model $b\to s\ell\bar\ell$ can be induced by leptoquark exchanging process at tree level.
The effective Hamiltonian to estimate the $b\to s\ell\bar\ell$ is given by 
\begin{align}
{\it H_{eff}} &= 
\frac12 \left(\frac{c^2_{\alpha_2}}{m_{A_2}^2} + \frac{s^2_{\alpha_2}}{m_{B_2}^2}\right)
f_{2\ell} f^\dag_{\ell3}
(\bar s\gamma_\mu P_R b) (\bar \ell \gamma^\mu P_L \ell)\nn\\
&- \frac{ g_{2\ell} g^\dag_{\ell3} }{2 m_{\delta_4}^2}
(\bar s\gamma_\mu P_L b) (\bar \ell \gamma^\mu P_L \ell)
+
\frac{ h_{2\ell} h^\dag_{\ell3} }{2 m_{\eta'_2}^2}
(\bar s\gamma_\mu P_L b) (\bar \ell \gamma^\mu P_R \ell),\label{eq:bsll}
\end{align}
where $m_{\delta_4}$ and $m_{\eta'_2}$ are respectively the mass of $\delta_{4/3}$ and $\eta'_{2/3}$.

Then, we decompose Eq.~(\ref{eq:bsll}) in terms of  the effective operators $O_9^{(')}$ and $O_{10}^{(')}$ defined by
 $O_{9(10)} = \bar s  \gamma_\mu P_L b \; \bar \ell \gamma^\mu (\gamma_5) \ell$,  $O'_{9(10)} = \bar s  \gamma_\mu P_R b \; \bar \ell \gamma^\mu (\gamma_5) \ell$, and 
their Wilson coefficients are found as
\begin{align}
C^{LQ,\ell}_9 &= -  \frac{1}{4 c_{\rm SM}} \left(\frac{g_{2\ell} g^\dag_{\ell3} }{m_{\delta_4}^2} 
-\frac{h_{2\ell} h^\dag_{\ell3} }{m_{\eta'_2}^2} \right), \quad
C^{LQ,\ell}_{10} = \frac{1}{4 c_{\rm SM}} 
 \left(\frac{g_{2\ell} g^\dag_{\ell3} }{m_{\delta_4}^2} 
+\frac{h_{2\ell} h^\dag_{\ell3} }{m_{\eta'_2}^2} \right)\,, \label{eq:C9C10}\\
C^{'LQ,\ell}_9 &=  \frac{1}{4 c_{\rm SM}} 
\left(\frac{c^2_{\alpha_2}}{m_{A_2}^2} + \frac{s^2_{\alpha_2}}{m_{B_2}^2}\right) f_{2\ell} f^\dag_{\ell3}, \quad
C^{'LQ,\ell}_{10} =- \frac{1}{4 c_{\rm SM}} 
\left(\frac{c^2_{\alpha_2}}{m_{A_2}^2} + \frac{s^2_{\alpha_2}}{m_{B_2}^2}\right)
f_{2\ell} f^\dag_{\ell3}\,, \label{eq:C9C10}
\end{align}
where $c_{\rm SM} = V_{tb} V^*_{ts} \alpha G_F/(\sqrt{2} \pi) $ is a scale factor from the SM effective Hamiltonian. 
$C^{LQ,\mu}_{10} $ also contributes to $B_s \to \mu^+ \mu^-$ whose
experimental data is consistent with the SM prediction.
In our numerical analysis we compare our results with best fit values of these Wilson coefficients 
obtained from global fit~\cite{Altmannshofer:2021qrr} that includes all the $B$ meson decay data. 
The best fit values for new physics contributions are 
\begin{align}
\label{BF1}
& C^\mu_9 = -0.91^{+0.18}_{-0.17} (-0.82^{+0.14}_{-0.14}) \quad \text{(for only $C_9$)}, \\
\label{BF2}
& C^\mu_{10} = +0.51^{+0.23}_{-0.24} (+0.56^{+0.12}_{-0.12}) \quad \text{(for only $C_{10}$)}, \\
\label{BF3}
& C^\mu_9 = C^\mu_{10} = -0.41^{+0.15}_{-0.15} (-0.06^{+0.11}_{-0.11}), \\
\label{BF4}
& C^\mu_9 = -C^\mu_{10} -0.65^{+0.12}_{-0.12} (-0.43^{+0.07}_{-0.07}), \\
\label{BF5}
& \{ C^\mu_9, C^\mu_{10} \} \simeq \{ -0.67, 0.24 \} \quad \text{( for $|C_9| \neq |C_{10}|$ for all rare $B$ decays)}, 
\end{align}
where values outside(inside) bracket are for $b \to s \mu \bar\mu$ observables only (all rare $B$ decays).
Among them cases, only the cases with $C^\mu_9$, $C^\mu_9 = -C^\mu_{10}$ and $|C_9| \neq |C_{10}|$ improve fit more than $5 \sigma$ compared to the SM case.
On the other hand $C_9^\mu = C_{10}^\mu$ case less improve the fit.

\subsection{LFVs and muon $g-2$} \label{lfv-lu}
Yukawa interactions associated with leptoquarks induce LFV decay of $\ell \to \ell' \gamma$ at one-loop level.
We then estimate the branching rations calculating relevant one-loop diagrams propagating leptoquarks.
Branching ratio of $\ell_a\to\ell_b\gamma$ 
is given by
\begin{align}
B(\ell_a\to\ell_b \gamma)
=
\frac{48\pi^3 C_{ab} \alpha_{\rm em}}{{\rm G_F^2} m_a^2 }(|(a_R)_{ab}|^2+|(a_L)_{ab}|^2),
\end{align}
where $m_{a(b)},\ (a(b)=1,2,3)$ is the mass for the initial(final) eigenstate of charged-lepton; $1\equiv e,2\equiv\mu,3\equiv\tau$, $(C_{21}, C_{31}, C_{32})=(1,0.1784, 0.1736)$.
$a_R$ is given by
\begin{align}
\label{eq:aR}
&(a_R)_{ab} \approx
  \frac{m_t}{(4 \pi)^2}h_{3a} \tilde h^\dagger_{b3} \int_0^1 dx\int_0^{1-x}dy   
  \left(  \frac{5}{x m_t^2 +(1-y) m_{\eta'_5}^2} 
  -
   \frac{2(1-x)}{x m_{\eta'_5}^2 +(1-y) m_{t}^2 }\right)\nn\\
&  +  \frac{m_c}{(4 \pi)^2}h_{2a} \tilde h^\dagger_{b2} \int_0^1 dx\int_0^{1-x}dy   
  \left(  \frac{5}{x m_c^2 +(1-y) m_{\eta'_5}^2} 
  -
   \frac{2(1-x)}{x m_{\eta'_5}^2 +(1-y) m_{c}^2 }\right)\nn\\
&-\frac{ g_{ia} g^\dag_{bi} m_a }{8(4\pi)^2}
\left(\frac{ s^2_{\alpha_1} }{m_{A_1}^2} + \frac{c^2_{\alpha_1} }{m_{B_1}^2}
- \frac{4 }{m_{\delta_4}^2}  \right) 
+\frac{3}{4(4\pi)^2}
\left(\frac{ h_{ia} h^\dagger_{bi} m_b }{m_{\eta'_5}^2} + \frac{ \tilde h^\dagger_{ia} \tilde h_{bi} m_a }{m_{\eta'_5}^2} \right),\end{align} 
where $m_t$ is the mass of top quark, and $a_L$ is obtained by $m_a\leftrightarrow m_b$ in the last line.
Clearly, the first term is dominant.
The current experimental upper bounds are given 
by~\cite{TheMEG:2016wtm, Adam:2013mnn}
  \begin{align}
  B(\mu\rightarrow e\gamma) &\leq4.2\times10^{-13},\quad 
  B(\tau\rightarrow \mu\gamma)\leq4.4\times10^{-8}, \quad  
  B(\tau\rightarrow e\gamma) \leq3.3\times10^{-8}~.
 \label{expLFV}
 \end{align}

 Our formula of the muon $g-2$ is given by
\begin{align}
\Delta a_\mu\approx -m_\mu [{(a_R)_{22}+(a_L)_{22}}].\label{damu}
\end{align}
We compare our value with the new result in Eq.~\eqref{eq:amu_new}.

\subsection{Neutral meson mixings $M-\overline{M}$}
Mixings between meson and anti-meson are induced by box diagrams propagating leptoquarks.
We have to estimate the mixings, since $M-\overline{M}$ mixings ($M=K_0(d\bar s),B_d(d\bar b),B_s(s\bar b), D_0(c\bar u)$), are precisely measured and there exist strong constraints.
 This estimation can be achieved by effective operators~\cite{Gabbiani:1996hi}, and each of their formulae 
is given by
\begin{align}
\Delta m_{M_d}&\approx
-\frac{m_{M_d} f_{M_d}^2}{6(4\pi)^2}
\sum_{a,a'=1}^3
\left(
{\rm Re}[f_{ja}f^\dagger_{ai} f_{ka'} f^\dagger_{a'\ell}]
\left[\sum_{q=1,2}\left(\frac{c^2_{\alpha_q}}{m^2_{A_q}} + \frac{s^2_{\alpha_q}}{m^2_{B_q}} 
+4 s_{\alpha_q}c_{\alpha_q} F_M(A_q,B_q)\right)\right]\right.\nn\\
&\left.+\frac14
{\rm Re}[g^\dagger_{aj}g_{ia} g^\dagger_{a'k} g_{\ell a'}]
\left[\frac{s^2_{\alpha_1}}{m^2_{A_1}} + \frac{c^2_{\alpha_1}}{m^2_{B_1}} 
-4 s_{\alpha_1}c_{\alpha_1} F_M(A_1,B_1) + \frac{4}{m^2_{\delta_4}}\right]\right.\nn\\
&\left.+\frac1{m^2_{\eta'_2}}
\left[{\rm Re}[h_{ja} h^\dagger_{ai} h_{ka'} h^\dagger_{a'\ell}] 
+
 {\rm Re}[\tilde h^\dagger_{ja}\tilde h_{ai} \tilde h^\dagger_{ka'} \tilde h_{a'\ell}]\right]
\right),
\end{align}

\begin{align}
\Delta m_{M_u}&\approx
-\frac{m_{M_u} f_{M_d}^u}{6(4\pi)^2}
\sum_{a,a'=1}^3
\left(
\frac14
{\rm Re}[g^\dagger_{aj}g_{ia} g^\dagger_{a'k} g_{\ell a'}]
\left[\sum_{q=1,2} q^2
\left(\frac{s^2_{\alpha_q}}{m^2_{A_q}} + \frac{c^2_{\alpha_q}}{m^2_{B_q}} 
-4 s_{\alpha_q}c_{\alpha_q} F_M(A_q,B_q)\right)\right]\right.\nn\\
&\left.+\frac1{m^2_{\eta'_5}}
\left[{\rm Re}[h_{ja} h^\dagger_{ai} h_{ka'} h^\dagger_{a'\ell}] 
+
 {\rm Re}[\tilde h^\dagger_{ja}\tilde h_{ai} \tilde h^\dagger_{ka'} \tilde h_{a'\ell}]\right]
\right),
\end{align}
where
$F_M(m_1,m_2)$ is given by
\begin{align}
F_M(m_1,m_2)&=\int_0^1dx \int_0^{1-x}dy\frac{1-x-y}{x m_1^2+ym_2^2}.
\end{align}
The current constraints are found as follows~\cite{Kumar:2020web, Zyla:2020zbs}:
\begin{align}
&\Delta m_{M_d}[(ijk\ell)=(1221)],\quad \Delta m_K \lesssim 3.48\times10^{-15}[{\rm GeV}],\label{eq:kk}\\
&\Delta m_{M_d}[(ijk\ell)=(1331)],\quad 
-1.85\times10^{-13} \ [{\rm GeV}]\lesssim\Delta m_{B_d} \lesssim 4.05\times10^{-14} \ [{\rm GeV}],\label{eq:bd}\\
&\Delta m_{M_d}[(ijk\ell)=(2332)],\quad 
-2.77\times10^{-12} \ [{\rm GeV}] \lesssim\Delta m_{B_s}  \lesssim 1.07\times10^{-12} \ [{\rm GeV}],\label{eq:bs}\\
&\Delta m_{M_u}[(ijk\ell)=(2112)],\quad \Delta m_D
 \lesssim 6.25\times10^{-15}[{\rm GeV}],\label{eq:uc}
\end{align}
where we apply the following values in our analysis: $f_K \approx 0.156$ GeV,
$f_{B_d(B_s)} \approx 0.191(0.274)$ GeV~\cite{DiLuzio:2017fdq, DiLuzio:2018wch},
$f_{D} \approx 0.212$ GeV,
$m_K \approx 0.498$ GeV, $m_{B_d(B_s)} \approx 5.280(5.367)$ GeV, and
$m_{D} \approx 1.865$ GeV.

\section{Numerical analysis \label{sec:numerical}}

In this section, we carry out numerical analysis scanning free parameters of the model searching for solutions 
to fit neutrino oscillation data and to explain $B$ anomalies and muon $g-2$, taking into account all the flavor constraints discussed above.
Firstly, we assume almost degenerate leptoquark masses $M_{LQ} \equiv m_{\delta4} = m_{\eta'_{2,5}} = m_{A_{1,2}}=m_{B_2}$ for simplicity; 
we only take $m_{B_2} = M_{LQ} \pm 10$ GeV to avoid too small $R_i$ in Eq.~\eqref{eq:R}. 
In fact degeneracy of masses for the components of $\eta^{(')}$ and $\Delta$  is motivated to suppress the oblique $S$- and $T$-parameters~\footnote{We thus do not explicitly consider these oblique parameters making them to be small by degenerate masses.}.
We then scan our parameters with the following ranges:
\begin{align}
& M_{LQ} \in [1\,, 5\,]\text{TeV},\quad |A_{12,23,13}| \in \left[10^{-13},\ 10^{-7}\,\right] \text{GeV}, \quad (\alpha_{1} ,\  \alpha_{2} ) \in [10^{-5}, 10^{-2}], \nn\\
&  | f_{ij} | \in [10^{-7}, 10^{-3}], \quad \{ |h_{1i}|, |h_{i1}| \} \in [10^{-7}, 10^{-3}], \quad  |h_{ab} (a, b \neq 1)| \in [10^{-5}, \sqrt{4 \pi}], \nn \\  
& \{ |\tilde h_{1i}|, |\tilde h_{i1}| \} \in [10^{-7}, 10^{-3}], \quad  |\tilde h_{ab} (a, b \neq 1)| \in [10^{-5}, \sqrt{4 \pi}].
\label{range_scanning}
\end{align}
Here we choose smaller values of Yukawa coupling associated with first generation to avoid strong flavor constraints. 
The Yukawa couplings $g_{ij}$ are obtained from Eq.~\eqref{eq:g} to fit neutrino oscillation data.
Note that we select smaller values for Yukawa coupling $f_{ij}$ so that $g_{ij}$ can be sizable to fit neutrino data.
As a result $C'^{LQ,\ell}_{9,10}$ is suppressed and we do not show these values explicitly.
Then we impose constraints from $BR(\ell \to \ell' \gamma)$ and $M$--$\bar M$ mixing discussed in previous section, 
and obtain the possible values of Wilson coefficient $C^{LQ,\mu}_{9,10}$ and $\Delta a_\mu$ for allowed parameter sets.

\begin{figure}[tb]
\begin{center}
\includegraphics[width=80mm]{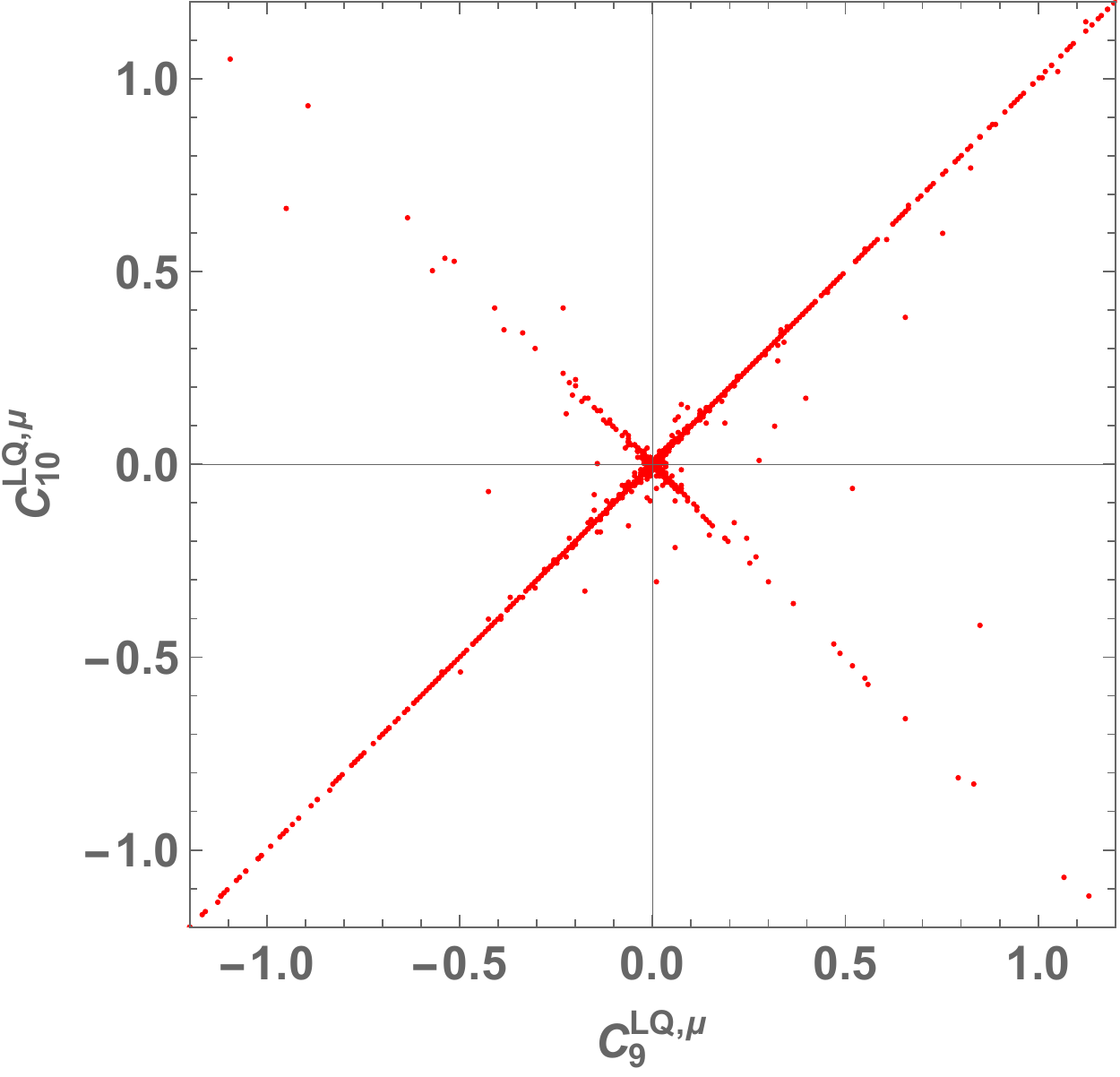}
\caption{$\Delta C_9^\mu$ and $\Delta C^\mu_{10}$ values for parameter sets satisfying neutrino data and all flavor constraints.
The shaded area indicates new muon $g-2$ value in Eq.~\eqref{eq:amu_new}. }
\label{fig:C9C10}
\end{center}
\end{figure}

In Fig.~\ref{fig:C9C10}, we show our results for $C^{LQ,\mu}_9$ and $C^{LQ, \mu}_{10}$ values where we impose $\Delta a_\mu > 0$ in addition to the other constraints.
We find that many parameter sets give $C^{LQ,\mu}_9 \sim \pm C^{LQ,\mu}_{10}$.
Note that we have  less points for same sign case $C^{LQ,\mu}_9 \sim C^{LQ,\mu}_{10}$ compared to opposite sign case since same sign contribution comes from Yukawa coupling $g$ that 
is more constrained by neutrino data.
Although some tunings of parameters are required, we can obtain the case of $|C^{LQ,\mu}_{10}| \neq |C^{LQ,\mu}_9|$ due to cancellation between diagrams.
Thus we have more freedom to obtain $\{C_9, C_{10} \}$ values thanks to the contributions from two different leptoquarks.
In principle we can obtain all the best fit values in ref.~\cite{Altmannshofer:2021qrr} summarized by Eqs.~\eqref{BF1}--\eqref{BF5} and that for $C^{LQ,\mu}_9 = - C^{LQ,\mu}_{10}$ case can be most easily realized.

\begin{figure}[tb]
\begin{center}
\includegraphics[width=80mm]{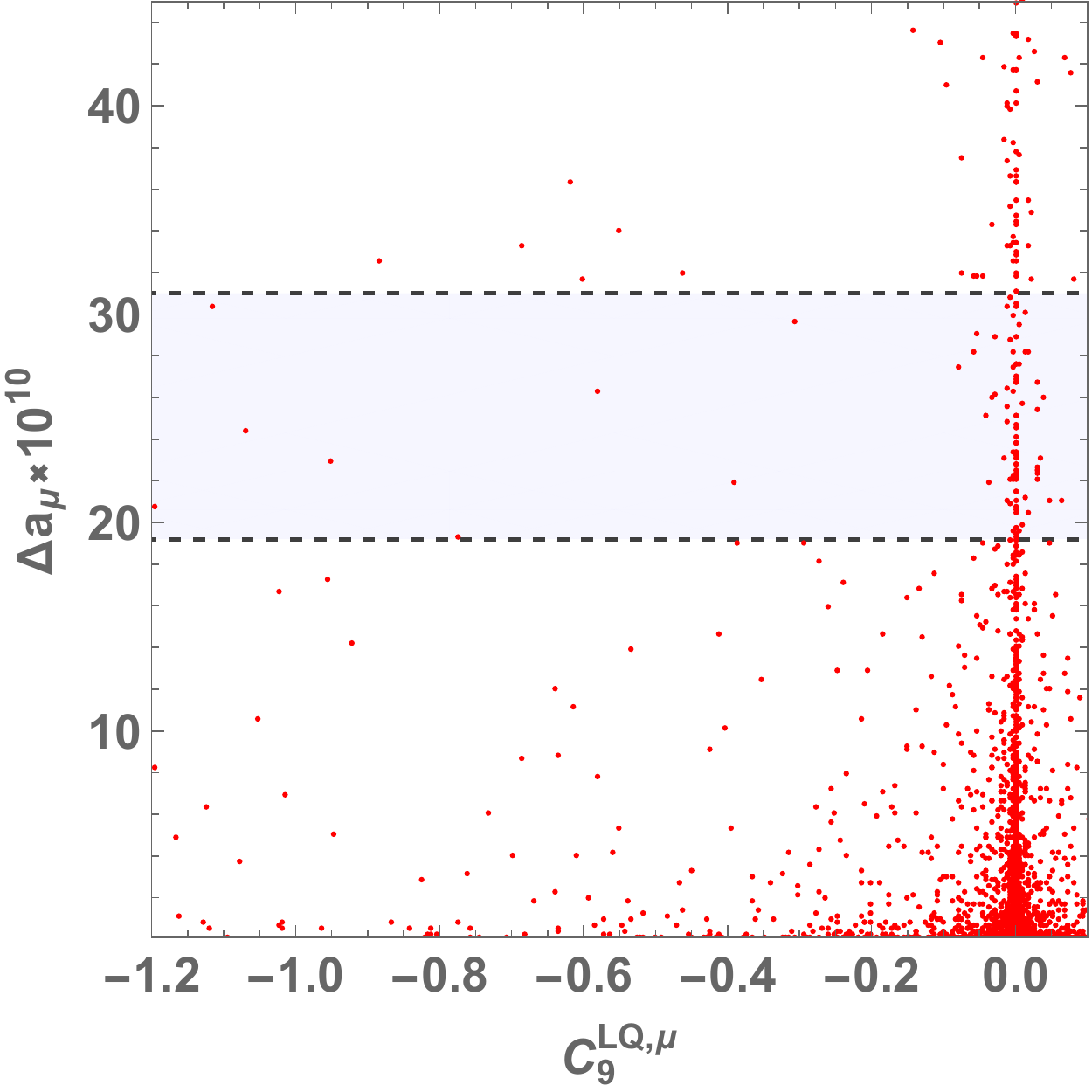}
\includegraphics[width=80mm]{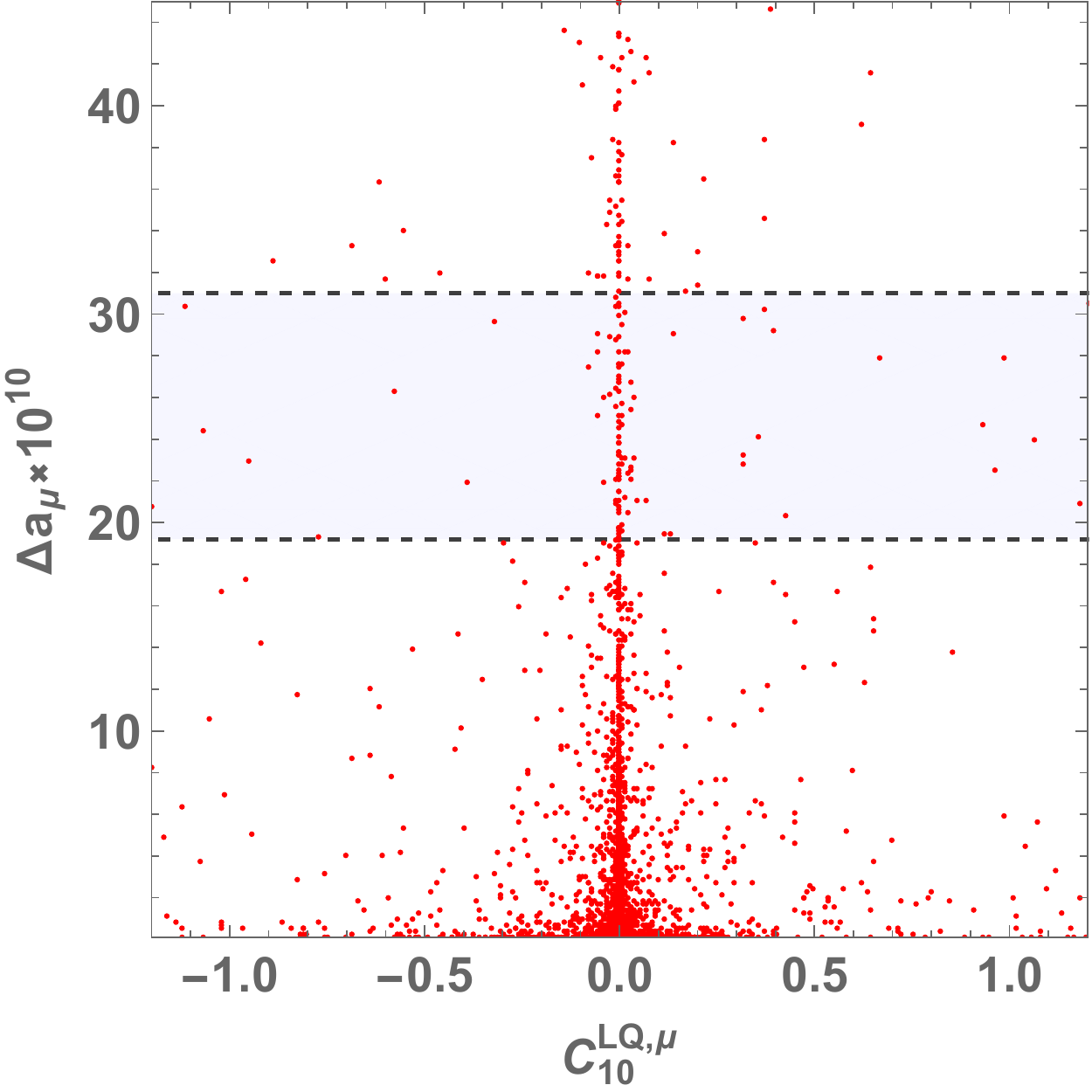}
\caption{$ C_{9(10)}^{LQ, \mu}$ and $\Delta a_\mu$ values for parameter sets satisfying neutrino data and all flavor constraints in the left(right) side figure. }
\label{fig:C9amu}
\end{center}
\end{figure}

In Fig.~\ref{fig:C9amu}, we show the values of $C^{LQ,\mu}_{9(10)}$ and $\Delta a_\mu$ that is compared with the new muon $g-2$ result Eq.~\eqref{eq:amu_new} in the left(right) figure.
As we see, it is possible to obtain the observed value of muon $g-2$ and $C^{LQ,\mu}_9 \sim -1$ at the same time; 
we also find both sign of $C_{10}^{LQ, \mu}$ is achieved. 
Note that sizable contribution is obtained due to enhancement by $m_t$ in Eq.~\eqref{eq:aR}.
Therefore we have parameter sets which explain both $b \to s \ell \bar\ell$ and muon $g-2$ anomalies.
Note that we don't find clear dependence on LQ masses and any values in $[1.0, 5.0]$ TeV are allowed.

\section{Conclusions}

We have discussed a model with three leptoquarks $\{\Delta, \eta, \eta' \}$ to explain $b \to s \ell \bar\ell$ anomalies, muon $g-2$ and neutrino masses, motivated by recent experimental results.
Active neutrino mass is generated at one-loop level via interactions among $\Delta$, $\eta$ and SM fermions.
The leptoquarks $\Delta$ and $\eta'$ contribute to Wilson coefficient $C_{9,10}^\ell$ at tree level.
Interestingly combination of two leptoquark contributions can provide a case of $|C_9^{LQ, \mu}| \neq |C_{10}^{LQ,\mu}|$ in contrast to one leptquark scenario.
In addition all three leptoquarks can provide contribution to muon $g-2$, and induce LFV decay processes and $M$--$\overline{M}$ mixing 
which are taken into account as constraints.
We then have carried out numerical analysis to explore $C_{9,10}^{LQ, \mu}$ and muon $g-2$ values when neutrino data and relevant flavor constraints are accommodated.
Our finding are as follows:
\begin{enumerate}
\item We can obtain sizable muon $g-2$ from $\eta'$ loop due to $m_t$ enhancement. 
Then we can fit the value of the new muon $g-2$ result shown in Eq.~\eqref{eq:amu_new}.

\item $C_9^{LQ, \mu} = - C_{10}^{LQ, \mu}$ case is preferred in our scenario but we can obtain $|C_9^{LQ, \mu}| \neq |C_{10}^{LQ, \mu}|$ case due to contributions from two leptoquarks.
It is thus possible to obtain the best fit values from global analysis not only for $C_9^{LQ, \mu} = - C_{10}^{LQ, \mu}$.
In addition we can accommodate explanation of $b \to s \ell \bar\ell$ and muon $g-2$ at the same time.

\item Leptoquark masses are not constrained if they are TeV scale. 
The collider experiments would constrain the mass range in future.

\end{enumerate}
 
Thus combination of leptoquarks is attractive scenario to explain $b \to s \ell \ell$ anomaly, muon $g-2$ and neutrino mass at the same time.

\section*{Acknowledgments}
The work of H.O. was supported by the Junior Research Group (JRG) Program at the Asia-Pacific Center for Theoretical
Physics (APCTP) through the Science and Technology Promotion Fund and Lottery Fund of the Korean Government and was supported by the Korean Local Governments-Gyeongsangbuk-do Province and Pohang City.
H.O. is sincerely grateful for all the KIAS members.


\end{document}